\documentclass[twocolumn,showpacs,preprintnumbers,amsmath,amssymb]{revtex4}

\usepackage{graphicx}
\usepackage{dcolumn}
\usepackage{bm}

\begin{document}

\preprint{}
\input{epsf.tex}

\epsfverbosetrue

\title{Collective Interactions in an Array of Atoms Coupled to a Nanophotonic Waveguide}
\author{Hashem Zoubi} 

\affiliation{Max-Planck Institute for the Physics of Complex Systems, Noethnitzer Str. 38, 01187 Dresden, Germany}

\date{23 October 2013}

\begin{abstract}
A lattice of trapped atoms strongly coupled to a one-dimensional nanophotonic waveguide is investigated in exploiting the concept of polariton as the
system natural eigenstate. We apply a bosonization procedure, which was
presented separately by P. W. Anderson and V. M. Agranovich, to transform
excitation spin-half operators into interacting bosons, and which shown here to confirm the hard-core boson model. We derive
polariton-polariton kinematic interactions and study them by solving the scattering
problem. In using the excitation-photon
detuning as a control parameter, we examine the regime in which polaritons
behave as weakly interacting photons, and propose the system for realizing superfluidity of photons. We implement the kinematic
interaction as a mechanism for nonlinear optical processes that
provide an observation tool for the system properties, e.g. the interaction
strength produces a blue shift in pump-probe experiments.
\end{abstract}


\maketitle

\section{Introduction}

The interest in light-matter interactions continue to be of big importance for
fundamental physics and applications. The localization of a Bose-Einstein Condensate (BEC) of
ultracold atoms between
optical cavity mirrors has been realized and the strong coupling regime observed
\cite{Brennecke,Colombe}. In the recent years nanophotonic waveguides \cite{Vetsch,Goban} seem to have more advantages in
manipulating, trapping and detecting neutral cold and ultracold atoms over the
other conventional traps \cite{Metcalf}, e.g. in using optical lattices
\cite{Jaksch,Greiner}. Tapered optical
nanofibers with radius smaller than the guided field wavelength give rise to
hybrid modes with significant part of their energy in the evanescent fields  surrounding
the fiber \cite{NayakA,NayakB}. The atoms are
trapped on an array outside and parallel to the fiber by the coupling to the evanescent fields of counter
propagating red and blue detuned beams, as have been demonstrated for cesium
atoms \cite{Vetsch}. Using magic wavelengths state-insensitive and compensated nanofiber trap have
been implemented \cite{Goban}. The atoms are shown to be efficiently interrogated with resonant green
light field sent through the nanofiber and observed through transmission and
reflection spectra \cite{Vetsch,Goban}. The collective enhancement effect is shown to allow
the atomic lattice to form high-quality cavity within the nanofiber, and
impurity atom that designated within the cavity can experience strongly
enhanced coherent coupling with fiber photons \cite{Chang}. Atomic forces and
optical scattering is also addressed and shown to lead to
self-organization of the atomic positions along the nanophotonic waveguide \cite{Cirac}.

On the other hand a BEC has been realized for ultracold atoms of a dilute boson gas \cite{Dalfovo}, and for cavity
polaritons in semiconductor microcavities \cite{{Kasprzak}}. Photons as bosons in principle can be condensate into BEC and behave as a
superfluid. Two main features are required to achieve these phenomena: first
the photons need to acquire an effective mass with a finite energy at zero wave
number that can be realized within a cavity, and second a mechanism for photon-photon
interactions which can be achieved by active material medium. Chiao \cite{Chiao} proposed
conventional nonlinear optical susceptibility as a mechanism for interacting
photons in two dimensional Fabry-Perot resonator and discussed possible
superfluidity of photons. A BEC of photons has been realized experimentally for  optical cavity filled with dye solution at room
temperature \cite{Klaers}.

In our previous work \cite{ZoubiA,ZoubiRev} the linear optical spectra was evaluated for a linear atomic lattice strongly coupled
to one dimensional propagating fiber
photons, where the excitations and
photons are coherently mixed to introduce polaritons as the real system
eigenstates \cite{Kavokin,ZoubiB}; and the atoms are considered to be of two-level
systems with spin-half statistics. For the case of a single excitation at
most to appear in the system no meaning of statistics and excitations can be
treated either as bosons or fermions; while for
two excitations and more, excitations at different sites behave as bosons and
on-site excitations behave as fermions. Namely electronic excitations in a
lattice of two-level systems have no defined quantum statistics and they are termed paulions, but it is desirable to work either with bosons or
fermions. Here we discard the fermionic picture, even though the Jordan-Wigner transformation is
an exact one from paulions into fermions in one dimensional
systems \cite{Altland}; as part of our concern implies collective
excitations to behave as dilute boson gas, we concentrate only in the bosonic picture.

Different bosonization scenarios are available in the literature. The Holestein-Primakof bosonization is not a good choice, as the spin operator is
represented by square-root of boson operators, the fact that leads to more
complexity \cite{Holstein}. The Schwinger transformation is also not
useful here as each paulion operator is represented in terms of two kinds of boson
operators \cite{Altland}. Agranovich et al. \cite{Agranovich} suggested an exact transformation from paulions into
bosons, in which each spin operator is represented by an infinite
power series of boson operators. In the limit of low density of excitations it is a good approximation to keep the lowest order terms of the
series, and this limit agrees with the ad hoc bosonization proposed by
P. W. Anderson \cite{Anderson}. The transformation forbids two excitations from being
localized on the same atom site and results in kinematic interactions \cite{ZoubiC}.

In the present paper we apply the above transformation to extract polariton-polariton interactions,
which we plan to introduce as a significant mechanism for nonlinear optical processes and many-body
effects. In using the excitation-photon detuning as a control parameter for
the strength of the interaction, we emphasize the limit in which polaritons weakly interact and behave as dilute boson gas. In this
limit we examine the regime in which interacting polaritons can be treated as
interacting photons, and we show that the present set-up has all the features for achieving BEC and superfluidity of photons. To get more quantitative understanding of the above kinematic
interaction we study the polariton-polariton scattering. In the center of mass frame such scattering can
be modeled as a scattering of polaritons from a defect in the lattice, and we
examine the validity of using the hard-core boson model in the present context
and that hold for any density of excitations. We use the kinematic interactions
as a source of nonlinearity for different optical processes \cite{Boyd}, e.g. pump-probe
experiments, that allow extracting the interaction strength and
atom-atom correlations, with optical bistability behavior.

The paper is organized as follows: in section 2 we present the concept of
polariton. The polariton-polariton kinematic interactions are
derived in section 3. The scattering problem of polaritons is studied in section 4. Section 5 is about nonlinear optical processes. Conclusions appear in section 6. The polaritons scattering off an atom impurity is included in the Appendix.

\section{Excitation-Photon Strong Coupling: Nanophotonic Polaritons}

We start by treating one dimensional lattice of atoms resonantly coupled to one
dimensional propagating photons. The system can be realized
for tapered nanofiber parallel to the atomic lattice as in figure (1). A pair of counter propagating red-detuned beams form
attractive optical lattice, and a pair of blue-detuned beams
form repulsive optical lattice. The interference of these evanescent fields
surrounding optical nanofiber provides an array of optical microtraps in which
the cold atoms loaded \cite{Vetsch,Goban}. The experiments can easily achieve
hundreds of
atomic sites, and larger number is expected in the future for smaller lattice
constant. The set-up properties justify the introduction of the
concept of polariton as
a natural excitation in the strong coupling regime \cite{ZoubiA}.

\begin{figure}
\centerline{\epsfxsize=5cm \epsfbox{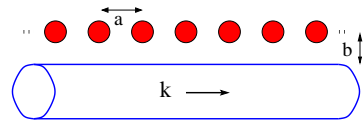}}
\caption{The atomic lattice of lattice constant $a$ is localized parallel to the
  nanofiber at distance $b$.}
\end{figure}

We consider a linear atomic lattice with one atom per site and lattice
constant $a$. The atoms are taken to be of two-level systems with atomic
transition $E_A$. The electronic excitation Hamiltonian is
\begin{equation}
H_A=E_A\sum_n B_n^{\dagger}B_n,
\end{equation}
where $B_n^{\dagger}$ and $B_n$ are the creation and annihilation operators of
an excitation at atom $n$, and the operators are of spin half. In the present
section as we deal with single excitation at most in the system we can assume the
operators to be of bosons with the commutation relation
$[B_n,B_m^{\dagger}]=\delta_{nm}$. From symmetry consideration, in the limit
of large number of lattice sites $N$, that is $N\gg 1$, where
the periodic boundary condition makes sense, we can use the Fourier transform
\begin{equation}
B_n=\frac{1}{\sqrt{N}}\sum_kB_ke^{-ikz_n},
\end{equation}
where the position of atom $n$ is
$z_n=an$, with $n=0,\pm1,\cdots,\pm M$, and $2M+1=N$. The wave number $k$ takes
the values $k=\frac{2\pi}{Na}l$, with $l=0,\pm1,\cdots,\pm M$. The Hamiltonian casts to
\begin{equation}
H_A=E_A\sum_k B_k^{\dagger}B_k.
\end{equation}
The excited atom have a finite life time,
e.g. the damping rate of the $D2$ transition of cesium atom at zero
temperature is $5.2\ MHz$. Later the excited atom damping rate is included
phenomenologically. Life times of electronic excitations in a lattice of atoms of finite size and in
which the interatomic distance can be larger or of the order of the transition
wavelength have been studied in \cite{ZoubiD}. It is shown that the life time
of excited atom is strongly affected by the existence of the other atoms in
the lattice.

We consider one dimensional propagating photons, and to concentrate in the
lowest fiber hybrid modes of $HE_{11}$. The photons are represented by the Hamiltonian
\begin{equation}
H_C=\sum_q\ E_C(q)\ a_q^{\dagger}a_q,
\end{equation}
where $a_q^{\dagger}$ and $a_q$ are the creation and annihilation operators of
a photon of mode $q$. In order to define the values of the wave number $q$, we
assume two parallel quantization mirrors, at the far sides of the
long nanofiber, and that are separated by a distance $L$. Hence, using period
boundary condition that allows propagating photons, the wave number $q$ is
quantized and takes the values $q=2\pi l/L$, where $l=0,\pm1,\pm2,\cdots,\pm\infty$. The photon dispersion can be given by
\begin{equation}
E_C(q)=\frac{\hbar c}{\sqrt{\epsilon}}\sqrt{q_0^2+q^2},
\end{equation}
where $\epsilon$ is an effective dielectric constant, and $q_0$ is the
confinement wave number that absorb all the complexity arouses in the fiber photon dispersion. The fiber photons have relatively long life time, but they
can be damped indirectly through their coupling to the atoms. The damping rate
of the cavity photons is contained later phenomenologically.

The excitation-photon coupling is taken in the electric dipole approximation
by $H_I=-\hat{\mbox{\boldmath$\mu$}}\cdot\hat{E}$, where the excitation
transition dipole operator is given by
$\hat{\mbox{\boldmath$\mu$}}=\mbox{\boldmath$\mu$}\sum_n\left(B_n+B_n^{\dagger}\right)$, and $\mbox{\boldmath$\mu$}$ is the transition dipole. The photon electric field operator is given by
\begin{equation}
\hat{E}(r,z)=i\sum_q\sqrt{\frac{E_C(q)}{2\epsilon_0V}}{\bf e}u(r)\left(a_qe^{-iqz}-a_q^{\dagger}e^{iqz}\right),
\end{equation}
where $V$ is the photon effective volume, ${\bf e}$ is the photon unit
vector polarization, and $u(r)$ is the photon mode function that can contain
all the complexity of the real fiber photon mode function.

The interaction Hamiltonian in the rotating wave approximation, and in the Schr$\ddot{o}$dinger picture, reads
\begin{eqnarray}
H_{AC}&=&-i\sum_{qn}\sqrt{\frac{E_C(q)}{2\epsilon_0V}}u(b)(\mbox{\boldmath$\mu$}\cdot{\bf
  e}) \nonumber \\
&\times&\left(a_qB_n^{\dagger}e^{-iqz_n}-B_na_q^{\dagger}e^{iqz_n}\right).
\end{eqnarray}
The electric field is taken at the atom positions with $u(b)$ the mode
function at the lattice position that is separated by $b$ from the fiber. We
consider here the case in which the fiber length is equal to the lattice
length. Using the
inverse of the above Fourier transform, and in using the known lattice identity
$\frac{1}{N}\sum_ne^{i(q-k)z_n}=\delta_{kq}$, at the limit of $N\gg 1$, we get
\begin{equation}
H_{AC}=\sum_{k}\left(g_k\ a_kB_k^{\dagger}+g_k^{\star}\ B_ka_k^{\dagger}\right),
\end{equation}
where we define the coupling parameter
\begin{equation}
g_k=-i\sqrt{\frac{E_C(k)N}{2\epsilon_0V}}u(b)(\mbox{\boldmath$\mu$}\cdot{\bf e}).
\end{equation}
Due to translational symmetry the interaction is between an excitation and a photon
with the same wave number.

The total Hamiltonian, as a results of exploiting the lattice symmetry, is separated for each $k$ and given by
\begin{equation} \label{HamBos}
H=\sum_{k}\left\{E_A\ B_k^{\dagger}B_k+E_C(k)\ a_k^{\dagger}a_k+g_k\ a_kB_k^{\dagger}+g_k^{\star}\ B_ka_k^{\dagger}\right\}.
\end{equation}
The Hamiltonian also includes the term $H_{dec}=\sum_{q\neq k}E_C(q)\
a_q^{\dagger}a_q$, that represents all the photons
which decouple to the excitations. For the case where the fiber length is exactly equal
to the lattice length, that is $L=aN$, the last term represents all
photons with half wave length smaller than the lattice constant, that is
$a>\lambda/2$ where $q=2\pi/\lambda$. Or for photons with wave numbers larger
than the Brillouin zone boundary, that is $q>k_B$ where $k_B=\pi/a$. As in
the following we interest in excitation-photon resonances only
at small wave numbers, with $k\ll k_B$, the last term photons are far off resonance with the
atomic transition and they fall outside the first Brillouin zone
boundary. Namely we have $E_C(k_B)> E_A$. Hence we drop this term as it
includes photons that are not involved in the dynamics.

On the other hand the
fiber can naturally provide propagating evanescent fields only for modes with
wavelengths larger than the tapered fiber radius, which is in the experiment
about $250\ nm$, and all modes with wavelengths smaller than the fiber radius
are concentrated inside the fiber, which is the case for the $HE_{11}$ modes
that considered in the experiments \cite{Warken}. As the lattice constant is about $500\ nm$, the number of effective photon
modes is not much larger than the number of atom sites, the fact that supports our representation.

In the strong coupling regime where the excitation and photon line widths are
smaller than the coupling parameter, the excitations and photons are
coherently mixed to give two polariton branches. The Hamiltonian is
diagonalized by using the upper and lower polariton operators
\begin{equation}
A_k^{\pm}=X_k^{\pm}\ B_k+Y_k^{\pm}\ a_k,
\end{equation}
which is a coherent superposition of excitations and photons. The mixing
amplitudes are defined by
\begin{equation}
X_k^{\pm}=\pm\sqrt{\frac{ D_k\mp\delta_k}{2 D_k}},\ \ \ Y_k^{\pm}=\frac{g_k}{\sqrt{2 D_k( D_k\mp\delta_k)}},
 \end{equation}
where $D_k=\sqrt{\delta_k^2+|g_k|^2}$, and the detuning is
$\delta_k=\frac{E_C(k)-E_A}{2}$. The polariton Hamiltonian reads
\begin{equation}
H_{pol}=\sum_{k,\nu}E_{\nu}(k)\ A_k^{\nu\dagger}A_k^{\nu},
\end{equation}
with the polariton dispersions
\begin{equation}
E_{\pm}(k)=\frac{E_C(k)+E_A}{2}\pm D_k.
\end{equation}
We present the results for some typical numbers. The transition energy is $E_A=1\ eV$, the lattice constant is $a=5000\ \AA$, the dielectric
constant $\epsilon=4$, the transition dipole is $\mu=2\ e\AA$, and the mode
function is estimated to be $u(b)=0.25$. We also use $V=SNa$
with the effective area $S=\pi a^2$. We have resonance
excitation-photon at $k=0$, where $q_0\approx 10^{-3}\ \AA^{-1}$, with
$E_C(0)=E_A$. The boundary of the Brillouin zone is at $k_B\approx
6.28\times 10^{-4}\ \AA^{-1}$ with photon energy of $E_C(k_B)\approx
1.18\ eV$. Hence the photon at the Brillouin zone boundary are far off
resonance with the atomic transition, where $E_C(k_B)-E_A\approx 0.18\ eV$, that justifies the neglect of photons with wave numbers beyond $k_B$.

We plot the
polariton eigen-energies for the two branches in figure (2.a) with their excitation and photon
fractions in figure (2.b). The excitation and photon are coherently mix and split to give two
polariton branches that are separated by the Rabi splitting at the intersection
point. In the present zero detuning case we have Rabi splitting of $2|g_0|$,
which is about $1.5\times 10^{-6}\ eV$ or $3.7\ GHz$. Around the excitation-photon intersection point, at $k=0$, the polariton
is half excitation and half photon. For large $k$ the lower branch becomes
excitation and the upper one photon.

\begin{figure}
\centerline{\epsfxsize=4.5cm \epsfbox{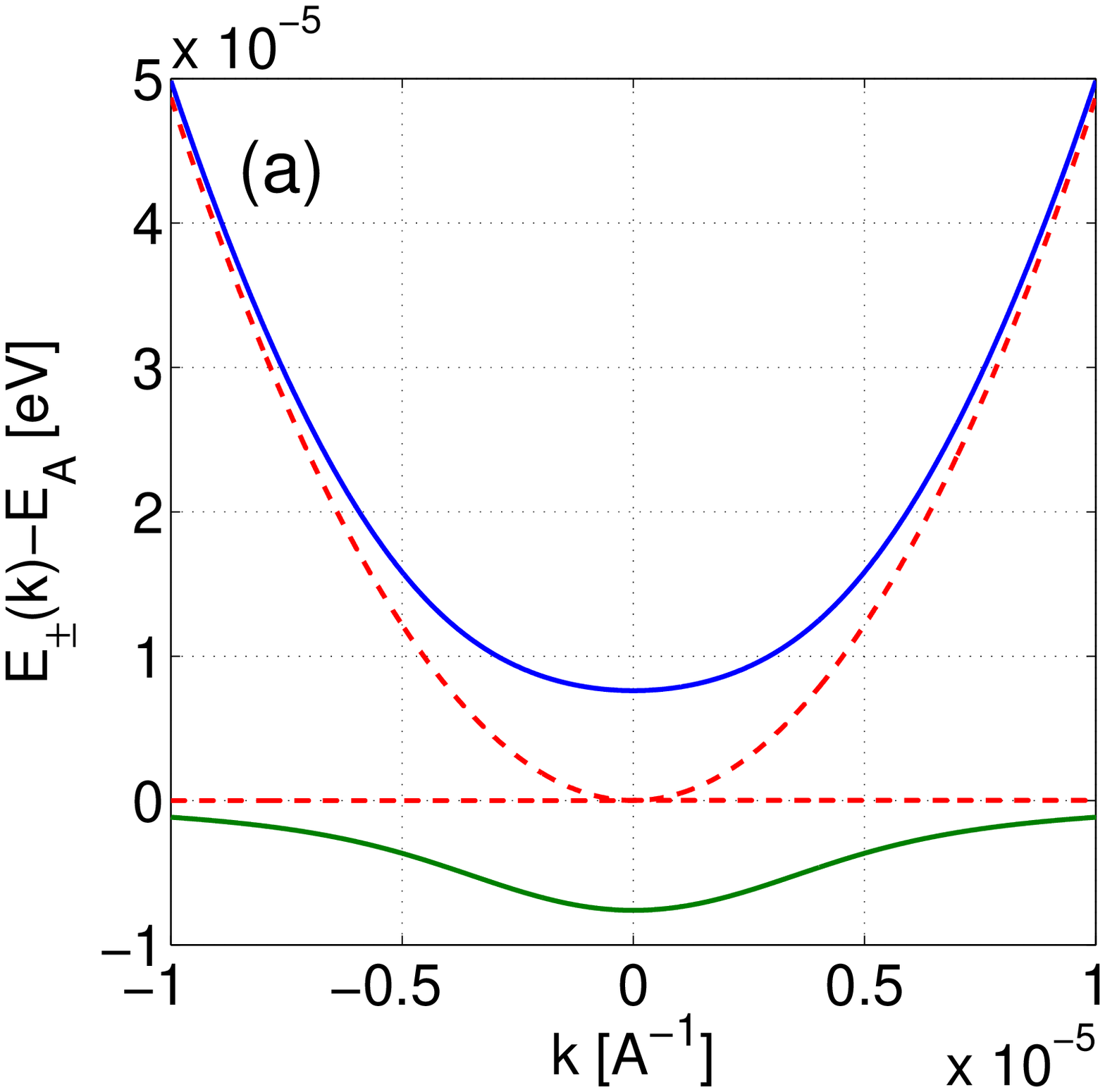}\epsfxsize=4.5cm \epsfbox{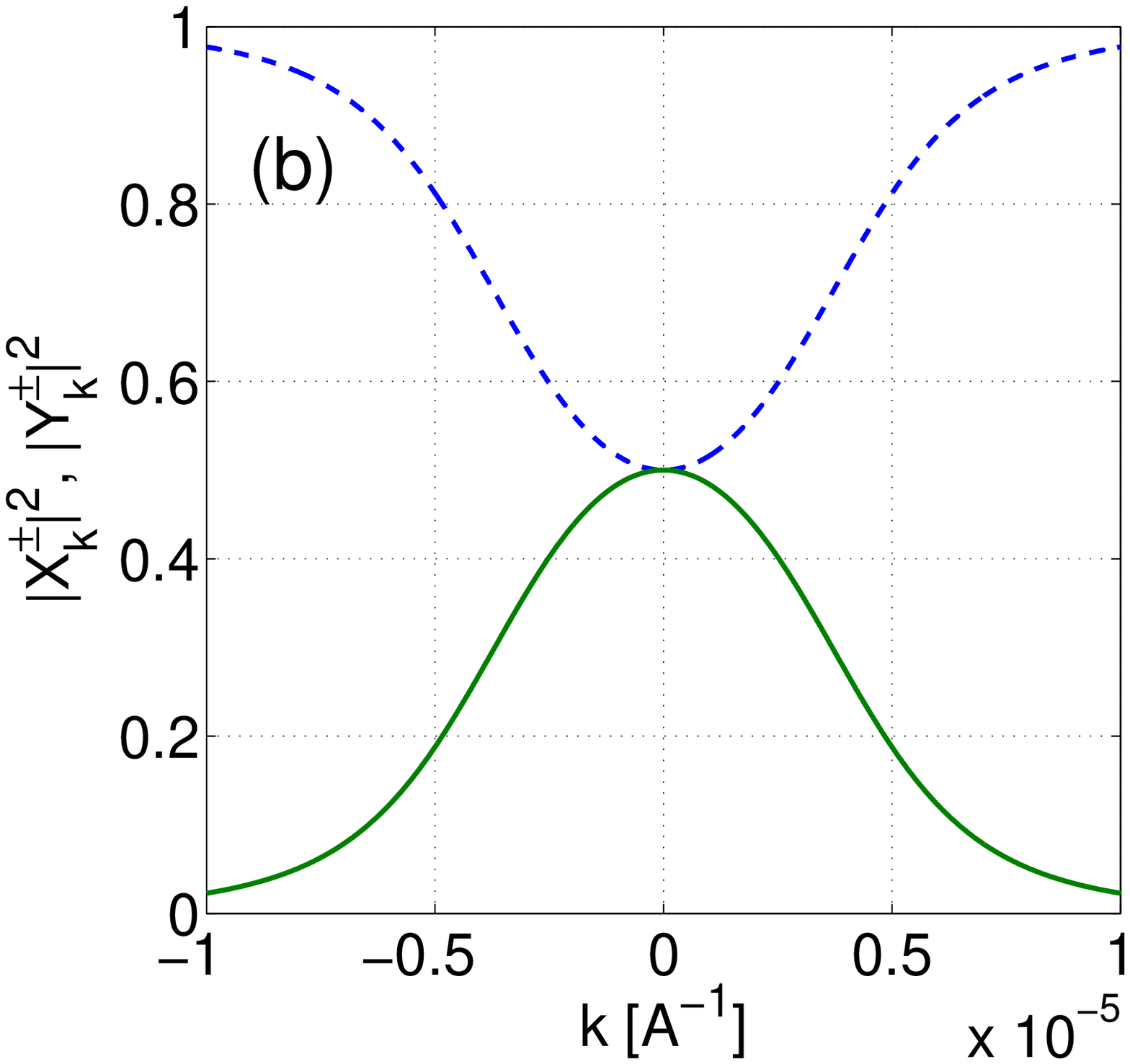}}
\caption{(a) The polariton relative energies $E_{\pm}(k)-E_A$ vs. $k$ for the upper and
  lower branches. The dashed lines are for the transition and the photon
  energies. (b) The excitation and photon fractions in the two polariton branches vs. $k$. In
  the lower branch the full line is for the photon fraction $|Y_k^{-}|^2$ and the
  dashed line for the excitation fraction $|X_k^{-}|^2$. In
  the upper branch the dashed line is for the photon fraction $|Y_k^{+}|^2$ and the
  full line for the excitation fraction $|X_k^{+}|^2$.}
\end{figure}

\section{Polariton-Polariton Kinematic Interactions}

Electronic excitations are treated here as two level systems where the
atomic operators are of spin-half. In the previous consideration we treated
single excitation to appear at most in the system and the operators are
assumed to be of bosons. For more excitations things
start to be different as each atom can be excited only once by absorbing a
photon. After the excitation the atom is saturated and the second photon will not
be absorbed by the same excited atom.

First let us present the properties of spin-half operators on a lattice,
$S_n^{\dagger}$ and $S_n$. They obey the fermi anti-commutation relation on the
same site, that is
\begin{equation}
S_nS_n^{\dagger}+S_n^{\dagger}S_n=1\ ,\ S_nS_n=S_n^{\dagger}S_n^{\dagger}=0,
\end{equation}
and the bose commutation relation between different sites, that is 
\begin{equation}
\left[S_n,S_m^{\dagger}\right]=\left[S_n,S_m\right]=\left[S_n^{\dagger},S_m^{\dagger}\right]=0,\
(n\neq m).
\end{equation}
Then spin-half operators on a lattice have mixed statistics. They are
fermions on-site and bosons among different sites, where they usually termed
paulions.

An exact transformation from paulions into
bosons was suggested by Agranovich et al. \cite{Agranovich}, in which each spin operator is represented in terms of infinite
power series of boson operators. At low density of excitations, when the number
of excitations is taken to be much smaller than the number of atoms in the
system, it is a good approximation to keep the lowest terms of the series. Hence
we have the transformation
\begin{equation}
S_n\rightarrow \left(1-B_n^{\dagger}B_n\right)B_n\ ,\ S_n^{\dagger}\rightarrow B_n^{\dagger}\left(1-B_n^{\dagger}B_n\right).
\end{equation}
The transformation results in the previous boson Hamiltonian (\ref{HamBos}), and in the additional kinematic interaction Hamiltonian, as it is due to quantum
statistics,
\begin{equation}
H_I=U\sum_nB_n^{\dagger}B_n^{\dagger}B_nB_n,
\end{equation}
where here $U=-E_A$. We neglect the interaction term results of the excitation-photon coupling, as
we are in the limit of $E_A\gg |g_k|$. The present bosonization transforms the
system from free paulions into interacting bosons. The interaction is on-site with
parameter $E_A$. The interaction is attractive, but the formation of bound
states is immaterial as it is of deep energy of $2E_A$ and no source avaliable
to absorb such energy. Hence the interaction can lead only to scattering among
the particles. In the hard-core boson model presented later we have the limit
$U\rightarrow\infty$, and the above transformation becomes
exact and hold for any density of excitations. The dynamical interactions due
to electrostatic forces among neutral atoms are much smaller than the
above kinematic ones at the present experimental interatomic distances.

In momentum space, by using the previous excitation operators, the interaction reads
\begin{equation}
H_I=\frac{U}{N}\sum_{kk'\bar{k}}B_{k'-\bar{k}}^{\dagger}B_{k+\bar{k}}^{\dagger}B_{k'}B_{k}.
\end{equation}
The next step is to write the interaction in terms of polariton operators, to get
\begin{equation}
H_I=\frac{U}{N}\sum_{kk'\bar{k}}\sum_{rsuv}\left(X_k^{r\star}X_{k'}^{s\star}X_{k+\bar{k}}^{u}X_{k'-\bar{k}}^{v}\right)A_{k'-\bar{k}}^{v\dagger}A_{k+\bar{k}}^{u\dagger}A_{k'}^sA_{k}^r.
\end{equation}
The Hamiltonian represents the interaction between two polaritons initially in
branches $r$ and $s$ with momenta $k$ and $k'$, and after the interaction in
branches $u$ and $v$ with momenta $k+\bar{k}$ and $k'-\bar{k}$, where they exchange
momentum $\bar{k}$ and can also change branches.

Let us concentrate here in the interaction among lower branch
polaritons. We assume the higher branch to be un-populated and we neglect the
possibility of the scattering of two lower polaritons with final states
in the upper branch. This process can be excluded due to conservation of
energy and momentum. We treat polaritons with small wave numbers and
around the minimum energy of the lower branch.  These polaritons can
be assumed to have a parabolic dispersion with a mass of the order of
the cavity photon effective mass. Hence we can drop here the branch index. As $X_k$ are smooth functions of $k$ for small $k$ we can neglect the
dependent of $X_k$ on $k$ and we use $X_0$. The interaction can be at any lattice site with the same probability of $1/N$. Finally we can write
\begin{equation}
H_I=\frac{a}{L}U|X_0|^4\sum_{kk'\bar{k}}A_{k'-\bar{k}}^{\dagger}A_{k+\bar{k}}^{\dagger}A_{k'}A_{k},
\end{equation}
where
$|X_0|^4=\frac{\left(\sqrt{\delta_0^2+|g_0|^2}+\delta_0\right)^2}{4\left(\delta_0^2+|g_0|^2\right)}$,
with $N=L/a$.

\begin{figure}
\centerline{\epsfxsize=6cm \epsfbox{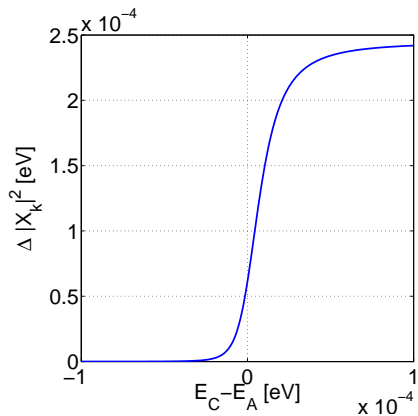}}
\caption{The scattering strength $\Delta |X_k|^4$ vs. the excitation-photon detuning $E_C(k)-E_A$, at $k=10^{-6}\ \AA^{-1}$.}
\end{figure}

In the hard core boson model the scattering length is exactly the potential
range, which is for the lattice case the lattice constant $a$. In the next
section we give detail calculation of the scattering problem that confirm
using the hard-core boson model. For large
wavelength polaritons, the
interaction strength $U$ can be represented in term of the scattering length
\cite{Dalibard}, by $U=\frac{4\pi\hbar^2}{m_{eff}a^2}$, where $m_{eff}$ is the polariton
effective mass that can be defined by
$\frac{1}{m_{eff}}=\frac{1}{\hbar^2}\frac{\partial^2 E_{\nu}(k)}{\partial
  k^2}$. Hence we get the effective interaction Hamiltonian
\begin{equation}
H_I=\Delta\sum_{rsuv}\sum_{kk'\bar{k}}\left(X_k^{r\star}X_{k'}^{s\star}X_{k+\bar{k}}^{u}X_{k'-\bar{k}}^{v}\right)A_{k'-\bar{k}}^{v\dagger}A_{k+\bar{k}}^{u\dagger}A_{k'}^sA_{k}^r,
\end{equation}
where $\Delta=\frac{4\pi\hbar^2}{m_{eff}aL}$. For the lower polariton branch around small wave numbers, $m_{eff}$ is of the
order of the photon effective mass. We can write
\begin{equation}
H_I=\Delta |X_0|^4\sum_{kk'\bar{k}}A_{k'-\bar{k}}^{\dagger}A_{k+\bar{k}}^{\dagger}A_{k'}A_{k}.
\end{equation}
In using the previous numbers, in figure (3) we plot the effective interaction as a
function of the excitation-photon detuning for a fixed wave number of
$k=10^{-6}\ \AA^{-1}$. We take the lattice length to be $L=1\ mm$. The photon
effective mass is about $m_{eff}c^2=4 eV$. It is clear that the interaction strength is strong around zero
detuning and decreased with
increasing the negative detuning.

We conclude here that the excitation-photon detuning serves as a significant
control parameter for the  strength of the kinematic interaction. By varying the
detuning from being positive where polaritons are mainly excitations into
positive where they are mainly photons, the polaritons switch from strongly
interacting quasi-particles into weakly interacting ones. For a fixed fiber
properties, the atomic excitation level can be changed in applying external fields.

We examine now the validity of considering the system of polaritons as a dilute degenerate
bose gas. The inter-polariton separation needs to be of the order of the de Broglie
thermal wavelength, that is defined by \cite{Dalfovo}
$\lambda_{dB}=\sqrt{\frac{2\pi\hbar^2}{m_{pol}k_BT}}$. For interacting polaritons the dilute Bose gas limit is for de Broglie wave
length much larger than the interaction range, that is $\lambda_{dB}\gg
a$. Using the previous numbers at resonance where $m_{pol}c^2\approx 4\ eV$ and at
room temperature, that is $k_BT\approx 25\ meV$, we get $\lambda_{dB}\approx
15\times 10^{3}\ \AA$, which is a bit larger than $a=5\times 10^{3}\ \AA$. Hence one needs either smaller lattice constant or to go to lower temperatures. At
temperature of $T=25\ K$ we have $k_BT\approx 2.5\ meV$ and we get $\lambda_{dB}\approx
5\times 10^{4}\ \AA$ which is one order of magnitude larger than $a$. 

The polariton effective mass for small wave numbers is of the order of the
cavity photon effective mass at small wave numbers. For small wave number
photons, or long wave length ones, in
the limit $q\ll q_0$, we have $E_C(q)\approx E_0+\frac{\hbar^2q^2}{2m_C}$, where $E_0=\frac{\hbar cq_0}{\sqrt{\epsilon}}$, and we defined the
confined photon effective mass $m_Cc^2=\hbar cq_0\sqrt{\epsilon}$. The
dispersion is parabolic with finite zero energy of $E_0$, the fact that has
big influence on the system properties that treated later.

This discussion is a step toward
achieving BEC of polaritons in such system, which is possible due to the
small effective mass and the finite minimum energy. Thermalization of
polaritons toward the minimum energy is a critical issue here and can be
achieved through polariton-polariton kinematic interactions. This topic needs
much more study that we leave for the future.

\section{Backward and Forward Scattering of Polaritons}

Let us now examine the scattering problem of two polaritons via the kinematic interaction, where the
scattering is only among the polariton excitation parts. We consider
analytically the scattering of lower branch polaritons with small wave
numbers. Here we neglect the scattering of lower polaritons into
upper ones as final states, that can be justified due to conservation of
energy and momentum, but the upper polaritons still appear as
intermediate scattering states. This case is of a single
channel scattering problem, and we can use well known results of scattering
theory. In the center of mass frame the problem is equivalent to the scattering of a polariton with reduced effective mass from the potential
results of the other
polariton, which is equivalent to the scattering of a polariton
from an effective potential that can be modeled as a defect in the lattice \cite{Baym,Newton}.

For the scattering of a polariton from a defect, we assume that single atom is missed in the ideal lattice as in figure (4). Namely we have
a localized vacancy at the origin site $n=0$. We add and subtract atomic Hamiltonian part at
the vacancy
site to get the above free polariton Hamiltonian $H_{pol}$ with additional impurity
Hamiltonian, that is we have $H=H_{pol}+H_{v}$ where $H_{v}=U\
B_0^{\dagger}B_0$. The vacancy appears as potential well of depth $E_A$ and width $a$. We will
not try to diagonalize the whole Hamiltonian, but we will treat the elastic
scattering problem.

\begin{figure}
\centerline{\epsfxsize=5.5cm \epsfbox{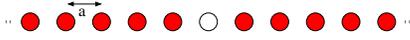}}
\caption{A lattice of lattice constant $a$ with a localized vacancy.}
\end{figure}

The polariton eigenstates obey $H_{pol}|\phi^{pol}_{{k}\nu}\rangle=E_{\nu}(k)|\phi^{pol}_{{k}\nu}\rangle$, and the whole system eigenstates obey $H|\psi\rangle=E|\psi\rangle$. The cavity photon eigenstate is defined by $a_{k}^{\dagger}|vac\rangle=|\phi_{k}^{ph}\rangle$, and the excitation eigenstate in quasi-momentum space is defined by $B_{k}^{\dagger}|vac\rangle=|\phi_{k}^{ex}\rangle$. The $\nu$ polariton branch eigenstate is defined by $A_{{k}\nu}^{\dagger}|vac\rangle=|\phi_{{k}\nu}^{pol}\rangle$, and in terms of excitation and photon states we get $|\phi_{{k}\pm}^{pol}\rangle=X_k^{\pm}\ |\phi_{k}^{ex}\rangle+Y_k^{\pm}\ |\phi_{k}^{ph}\rangle$. In terms of real lattice space, we get $|\phi_{{k}\pm}^{pol}\rangle=\frac{X_k^{\pm}}{\sqrt{N}}\sum_ne^{ikz_n}\ |\phi_n^{ex}\rangle+Y_k^{\pm}\ |\phi_{k}^{ph}\rangle$. A general state $|\psi\rangle$ can be expanded as $|\psi\rangle=\sum_{k}\psi_{k}^{ph}\ |\phi_{k}^{ph}\rangle+\sum_m\psi_m^{ex}\ |\phi_m^{ex}\rangle$. 

In the scattering problem the incoming polariton is prepared very far from the
impurity in the unperturbed state $|\phi_{{k}\nu}^{pol}\rangle$ and  the
scattered polariton is observed very far from the impurity. Hence, the whole
system eigenstates $|\psi\rangle$ needs to obey the boundary condition
$|\psi\rangle\rightarrow|\phi_{{k}\nu}^{pol}\rangle$ as $H_{v}\rightarrow
0$. The required solution is given by the Schwinger-Lippmann equation \cite{Newton}
\begin{equation} 
|\psi\rangle=|\phi_{{k}\nu}^{pol}\rangle+G_0H_{v}|\psi\rangle, 
\end{equation} 
where 
\begin{equation} \label{Green} 
G_0=\lim_{\eta\rightarrow 0_+}\left(\frac{1}{E-H_0+i\eta}\right). 
\end{equation} 
The sign of $+i\eta$ is chosen in such a way to ensure that the scattered
polariton propagates far from the impurity. 

We multiply the equation from the left by the lattice excitation eigenstate
$\langle\phi_n^{ex}|$, and insert the identity operator,
$\sum_m|\phi_m^{ex}\rangle\langle\phi_m^{ex}|=\hat{\bf 1}$, between $G_0$ and $H_v$, to get 
\begin{equation} 
\langle\phi_n^{ex}|\psi\rangle=\langle\phi_n^{ex}|\phi^{pol}_{{k}\nu}\rangle+\sum_m\langle\phi_n^{ex}|G_0|\phi_m^{ex}\rangle\langle\phi_m^{ex}|H_v|\psi\rangle. 
\end{equation} 

Using the above definitions, we get the scattered polariton excitation part amplitude by 
\begin{equation} 
\psi_n^{ex}=\frac{X_k^{\nu}}{\sqrt{N}}e^{ikz_n}+\frac{U}{N}\sum_{{k'}\mu}\lim_{\eta\rightarrow 0_+}\frac{|X_{k'}^{\mu}|^2\ e^{i{k'z_n}}}{E_{\nu}(k)-E_{\mu}(k')+i\eta}\ \psi_0^{ex}, 
\end{equation} 
where due to energy conservation we replaced the energy $E$ by the initial
energy $E_{\nu}(k)$. For the impurity site amplitude we have 
\begin{equation} 
\psi_0^{ex}=\frac{X_k^{\nu}}{\sqrt{N}}\left\{1-\frac{U}{N}\sum_{{k'}{\mu}}\lim_{\eta\rightarrow 0_+}\frac{|X_{k'}^{\mu}|^2}{E_{\nu}(k)-E_{\mu}(k')+i\eta}\right\}^{-1}. 
\end{equation} 
We thus obtain 
\begin{equation} 
\psi_n^{ex}=\frac{X_k^{\nu}}{\sqrt{N}}\left\{e^{i{kz}_n}-\frac{UI_{O}}{1+UI_{S}}\right\}. 
\end{equation} 
For simplicity, at large number of lattice sites,
the $k$-space can be assumed continuous and the sums converted to the
integral $\frac{1}{N}\sum_{k'}\rightarrow \frac{a}{2\pi}\int dk'$, we get 
\begin{eqnarray} 
I_{O}&=&\sum_{\mu}\frac{aU}{2\pi}\int dk'\lim_{\eta\rightarrow 0_+}\frac{|X_{k'}^{\mu}|^2\ e^{i{k'z}_n}}{E_{\nu}(k)-E_{\mu}(k')+i\eta}, \nonumber \\ 
I_{S}&=&\sum_{\mu}\frac{aU}{2\pi}\int dk'\lim_{\eta\rightarrow 0_+}\frac{|X_{k'}^{\mu}|^2}{E_{\nu}(k)-E_{\mu}(k')+i\eta}. \nonumber \\ 
\end{eqnarray} 

We calculate here analytically the scattering case with small wave number
polaritons, where one
can approximate the dispersion to be parabolic around $k\approx 0$, where we
have $E_{-}(k)\approx E_{-}(0)+\frac{\hbar^2k^2}{2m_{pol}}$, and we defined the polariton effective mass around $k\approx 0$. In the case
of excitation-photon close to resonance it is of the order of the cavity
photon effective mass. We also neglect the contribution of the upper branch
polaritons as intermediate states in the scattering. They are of higher
energy and their excitation fraction become smaller and smaller with
increasing $k$. Here the integrals cast into
\begin{eqnarray} 
I_{O}&=&\frac{aU}{2\pi}\frac{2m_{pol}}{\hbar^2}\int dk'\lim_{\eta\rightarrow 0_+}\frac{|X_{k'}^{-}|^2\ e^{i{k'z}_n}}{k^2-{k'}^2+i\eta}, \nonumber \\ 
I_{S}&=&\frac{aU}{2\pi}\frac{2m_{pol}}{\hbar^2}\int dk'\lim_{\eta\rightarrow 0_+}\frac{|X_{k'}^{-}|^2}{k^2-{k'}^2+i\eta}. \nonumber \\ 
\end{eqnarray} 
 The calculus of residue yields
\begin{equation} 
I_{O}=iU\frac{am_{pol}}{\hbar^2k}|X_{k}^{-}|^2\ e^{-i{kz}_n},\ \ \ I_{S}=iU\frac{am_{pol}}{\hbar^2k}|X_{k}^{-}|^2.
\end{equation} 
We write
\begin{equation} 
\psi_n^{ex}=\frac{X_k^{\nu}}{\sqrt{N}}\left\{e^{i{kz}_n}+f(k)e^{-ikz_n}\right\},
\end{equation} 
where we defined the scattering amplitude by
\begin{equation}
f(k)=\frac{i\frac{U}{\Lambda}\left(\frac{\pi^2}{2ka}\right)|X_{k}^{-}|^2}{1-i\frac{U}{\Lambda}\left(\frac{\pi^2}{2ka}\right)|X_{k}^{-}|^2},
\end{equation} 
and we defined the energy $\Lambda=\frac{\hbar^2\pi^2}{2a^2m_{pol}}$. The square absolute value is given by
\begin{equation}
|f(k)|^2=\frac{\frac{U^2}{\Lambda^2}\left(\frac{\pi^2}{2ka}\right)^2|X_{k}^{-}|^4}{1+\frac{U^2}{\Lambda^2}\left(\frac{\pi^2}{2ka}\right)^2|X_{k}^{-}|^4}.
\end{equation}

As $U\gg\Lambda$, the excitation part scattering of small wave vector lower polaritons is complete
with $|f(k)|^2\approx 1$. We conclude that the polariton reflection is $|X_{k}^{-}|^2$ and the
polariton transmission is $|Y_{k}^{-}|^2$. One need to worry about the case of
large positive detuning as the parabolic assumption breaks down and numerical
calculation is needed, but the conclusion is the same. We obtain
\begin{equation} 
\psi_n^{ex}=\frac{X_k^{\nu}}{\sqrt{N}}\left\{e^{i{kz}_n}-e^{-ikz_n}\right\}.
\end{equation}
We showed that the reflection amplitude is exactly the
excitation amplitude and the transmission amplitude is the photon amplitude for
the scatterer polaritons. The results of this limit is exactly that of the
scattering from a hard core potential of radius $a$, where it is in the limit of
$U\rightarrow \infty$ for the defect potential, and that verifies the use
of the hard-core boson model in the present context. In the appendix we
implement the above result to the scattering of polaritons from an impurity
consists of an atom of a different type in
the lattice.

The above conclusion can be easily generalized to the scattering of any pair
of polaritons with any wave numbers and in any branches. Careful inspection of
the polariton excitation part scattering amplitude reveals the fact that the
polariton excitation parts of two polaritons can be only back scattered, in
the one dimensional case. This fact is due to the large value of $U$ relative to the other system
parameters, and much more strongly, in the hard-core boson model this conclusion
is exact. Hence if we consider two incident polaritons with wave numbers $k_1$
and $k_2$ and in polariton branches $r$ and $s$, respectively, the scattered
polaritons are with wave numbers $k_1^{\prime}$ and $k_2^{\prime}$, and in the
branches $r^{\prime}$ and $s^{\prime}$, where the elastic scattering respects
conservation of energy and momentum
\begin{equation}
E_r(k_1)+E_s(k_2)=E_{r^{\prime}}(k_1^{\prime})+E_{s^{\prime}}(k_2^{\prime}),\ |k_1+k_2|=|k_1^{\prime}+k_2^{\prime}|.
\end{equation}
The scattering is represented by the scattering matrix
\begin{equation}
\left( \begin{array}{cc}
A_{k_1^{\prime}}^{r^{\prime}} \\ A_{k_2^{\prime}}^{s^{\prime}} 
\end{array} \right) = \left( \begin{array}{ccc}
X_{k_1}^r & Y^s_{k_2} \\ 
Y^r_{k_1} & X_{k_2}^s 
\end{array} \right) 
\left( \begin{array}{cc}
A_{k_1}^{r} \\ A_{k_2}^{s} \end{array} \right).
\end{equation}
The scattering process is plotted in figure (5).

For the lower branch polaritons around small wave numbers the dispersion is
parabolic with an effective mass, hence the conservation of energy condition becomes
$k_1^2+k_2^2=k_1^{\prime 2}+k_2^{\prime 2}$. We have only two possible scattering
channels: forward scattering with $k_1=k_2^{\prime}$ and $k_2=k_1^{\prime}$,
or backward scattering with $k_1=k_1^{\prime}$ and $k_2=k_2^{\prime}$. The
process can lead to the formation of entangled polaritons discriminated by
their wave numbers. For two incident polaritons with $k_1$ and $k_2$ in the lower
branch, we get the final entangled state $|Ent(k_1,k_2)\rangle=Y_{k_1}Y_{k_2}|k_1,k_2\rangle+X_{k_1}X_{k_2}|k_2,k_1\rangle$.

\begin{figure}
\centerline{\epsfxsize=5cm \epsfbox{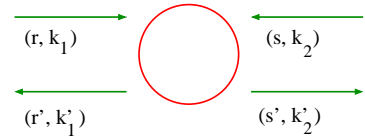}}
\caption{Schematic plot of two incident polaritons $(r,k_1)$ and $(s,k_2)$,
  scattered into final polaritons $(r^{\prime},k_1^{\prime})$ and $(s^{\prime},k_2^{\prime})$.}
\end{figure}

Polaritons in the case of large negative detuning are weakly interact and can
form a superfluid. In this limit as the polaritons are mainly photons we can
treat them as interacting photons and
there is a place to examine the superfluidity of photons. The main reasons
that allow
us to consider the superfluidity of photons are: First as polaritons include very small component of excitation they become interacting particles. Second
we dealing here with confined photons and they have effective mass and
non-zero energy at zero wave number. The only point to worry about is the
finite life time of polaritons which is due to the life time of the
photons and the excited atoms.

These features are important toward achieving the superfluidity of
photons. The finite
minimum energy needs to be larger than the thermal energy as then the
disappearance of photons within the cavity can be negligible and one get finite photon chemical
potential, which is important for defining conserved number of photons in
the system. Using the Bogoliubov theory for a dilute Bose
gas, the elementary excitation spectrum of the superfluid of photons is linear
with sound wave excitations. The sound waves are found to be
\cite{Dalfovo,Chiao} $v_s=\sqrt{\frac{N_0}{N}\frac{2\bar{U}}{m_{pol}c^2}}c$, where here $N_0$ is the number of condensate polaritons, $N$ the total
number of lattice sites, and $\bar{U}$ is an effective polariton-polariton
potential that can be extracted from the above scattering amplitudes. We use
the previous effective potential of $\bar{U}=\Delta|X_k|^4$.

\section{Polariton Nonlinear Optical Processes}

We aim now to apply the polariton-polariton kinematic interaction as a
mechanism for nonlinear optical processes \cite{Boyd}, and that provide a
direct observation tool for the strength of the above kinematic interaction. We start by formal discussion in
deriving the mean field equations of motion and including the damping rates
phenomenologically with the coupling to external fields, and later we concentrate
in few specific cases. The interacting polariton Hamiltonian reads
\begin{equation}
H=\sum_{k,r}E_{r}(k)\
A_k^{r\dagger}A_k^{r}+\sum_{rsuv}\sum_{k,k',\bar{k}}U^{rsuv}_{k,k',\bar{k}}\ A_{k'-\bar{k}}^{v\dagger}A_{k+\bar{k}}^{u\dagger}A_{k'}^sA_{k}^r,
\end{equation}
where $U^{rsuv}_{k,k',\bar{k}}=\Delta\
\left(X_k^{r\star}X_{k'}^{s\star}X_{k+\bar{k}}^{u}X_{k'-\bar{k}}^{v}\right)$.

The equation of motion for the polariton operator, in the Heisenberg picture,
is given by $i\hbar\frac{d}{dt}A_{k''}^{w}=[A_{k''}^{w},H]$, that yields
\begin{eqnarray}
i\frac{d}{dt}A_{k''}^{w}&=&\left(\Omega_{k''}^{w}-i\Gamma_{k''}^{w}\right)A_{k''}^{w}+F_{k''}^{w}
\\ \nonumber
&+&\sum_{rsu}\sum_{k,k'}V^{rsuw}_{k,k',k''}\ A_{k+k'-k''}^{u\dagger}A_{k'}^sA_{k}^r.
\end{eqnarray}
We used $E_{r}(k)=\hbar\Omega_{k}^{r}$, and defined $\hbar
V^{rsuw}_{k,k',k''}=U^{rsuw}_{k,k',k'-k''}+U^{rswu}_{k,k',k''-k}$. The polariton damping rate $\Gamma_{k}^{r}$, of branch $r$ and wave number
$k$, is included phenomenologically. It is a good approximation when the
thermal energy is much smaller than the polariton energy. The damping is due to
the life times of both the fiber photons and the excited atoms, where we can
write
\begin{equation}
\Gamma_{k}^{r}=\frac{\Gamma_A}{2}|X_{k}^{r}|^2+\frac{\Gamma_C}{2}|Y_{k}^{r}|^2,
\end{equation}
with $\Gamma_A$ is the excited atom damping rate, and $\Gamma_C$ is the fiber
photon damping rate.

Furthermore, we include in the equation of motion
external pump of polaritons, $F_{k}^{r}$, at branch $r$ and wave number
$k$. We consider external pumps at the two far edges of the fiber, by
assuming effective mirrors at the far sides of the fiber that couple the fiber
photon to external pump fields. Namely we have the Hamiltonian
$H_{ex}=\hbar\gamma
\sum_k\left(a_{k}\mathcal{F}^{\ast}_k+a_{k}^{\dagger}\mathcal{F}_k\right)$, where
$\gamma$ is the coupling parameter at the mirrors, and $\mathcal{F}_k$ is the
external classical field of wave number $k$. In
term of polaritons, by using the inverse transformation
$a_{k}=\sum_rY_{k}^{r\ast}A_k^{r}$, we obtain
$H_{ex}=\hbar\sum_{k,r}\left(F^{r\star}_k A_k^{r}+F^{r}_k
  A_k^{r\dagger}\right)$, where we defined $F^{r}_k=\gamma Y_{k}^{r\ast}\mathcal{F}_k$. This Hamiltonian
leads to the external field term in the above equation of motion.

We concentrate mainly in lower branch polaritons, and we drop the branch
indices. Then we have
\begin{eqnarray}
i\frac{d}{dt}A_{k''}&=&\left(\Omega_{k''}-i\Gamma_{k''}\right)A_{k''}
\nonumber \\
&+&\sum_{k,k'}V_{k,k',k''}\ A_{k+k'-k''}^{\dagger}A_{k'}A_{k}+F_{k''}.
\end{eqnarray}
We consider only polaritons of two fixed wave numbers, $k_1$ and $k_2$. As we
discussed before due to conservation of energy and momentum, we have only two
possible channels: (i) forward scattering $(k_1,k_2)\rightarrow (k_1,k_2)$, and
(ii) backward scattering $(k_1,k_2)\rightarrow (k_2,k_1)$. We obtain the two
equations of motion
\begin{eqnarray}
i\frac{d}{dt}A_{k_1}&=&\left(\Omega_{k_1}-i\Gamma_{k_1}\right)A_{k_1}+V_{k_1,k_2}\
A_{k_2}^{\dagger}A_{k_2}A_{k_1}+F_{k_1}, \nonumber \\
i\frac{d}{dt}A_{k_2}&=&\left(\Omega_{k_2}-i\Gamma_{k_2}\right)A_{k_2}+V_{k_1,k_2}\
A_{k_1}^{\dagger}A_{k_1}A_{k_2}+F_{k_2}, \nonumber \\
\end{eqnarray}
where $\hbar V_{k_1,k_2}=2\Delta|X_{k_1}|^2|X_{k_2}|^2$.

The two external fields are oscillating with frequencies $\omega_1$ and
$\omega_2$. Namely
\begin{equation}
F_{k_1}=\tilde{F}_{k_1}e^{-i\omega_1t},\ F_{k_2}=\tilde{F}_{k_2}e^{-i\omega_2t},
\end{equation}
and the polaritons oscillate with the same frequencies
\begin{equation}
A_{k_1}=\tilde{A}_{k_1}e^{-i\omega_1t},\ A_{k_2}=\tilde{A}_{k_2}e^{-i\omega_2t},
\end{equation}
which yield equations of motion in rotating frames
\begin{eqnarray}
i\frac{d}{dt}\tilde{A}_{k_1}&=&\left(\Omega_{k_1}-\omega_1-i\Gamma_{k_1}\right)\tilde{A}_{k_1} \nonumber \\
&+&V_{k_1,k_2}\
\tilde{A}_{k_2}^{\dagger}\tilde{A}_{k_2}\tilde{A}_{k_1}+\tilde{F}_{k_1}, \nonumber \\
i\frac{d}{dt}\tilde{A}_{k_2}&=&\left(\Omega_{k_2}-\omega_2-i\Gamma_{k_2}\right)\tilde{A}_{k_2} \nonumber \\
&+&V_{k_1,k_2}\
\tilde{A}_{k_1}^{\dagger}\tilde{A}_{k_1}\tilde{A}_{k_2}+\tilde{F}_{k_2}.
\end{eqnarray}

At this point we apply the mean field theory by defining the expectation value
of the polariton operators by
$\tilde{\mathcal{A}}_{k_1}=\langle\tilde{A}_{k_1}\rangle$, and
$\tilde{\mathcal{A}}_{k_2}=\langle\tilde{A}_{k_2}\rangle$. Moreover we apply the factorization approximation, in the mean field,
$\langle\tilde{A}_{k_2}^{\dagger}\tilde{A}_{k_2}\tilde{A}_{k_1}\rangle=\langle\tilde{A}_{k_2}^{\dagger}\tilde{A}_{k_2}\rangle\langle\tilde{A}_{k_1}\rangle$
and
$\langle\tilde{A}_{k_1}^{\dagger}\tilde{A}_{k_1}\tilde{A}_{k_2}\rangle=\langle\tilde{A}_{k_1}^{\dagger}\tilde{A}_{k_1}\rangle\langle\tilde{A}_{k_2}\rangle$. We
can write
$\langle\tilde{A}_{k_2}^{\dagger}\tilde{A}_{k_2}\tilde{A}_{k_1}\rangle=\mathcal{N}_{k_2}\tilde{\mathcal{A}}_{k_1}$
and
$\langle\tilde{A}_{k_1}^{\dagger}\tilde{A}_{k_1}\tilde{A}_{k_2}\rangle=\mathcal{N}_{k_1}\tilde{\mathcal{A}}_{k_2}$,
where we defined the expectation value of the number operators by
$\mathcal{N}_{k_1}=\langle\tilde{A}_{k_1}^{\dagger}\tilde{A}_{k_1}\rangle$ and
$\mathcal{N}_{k_2}=\langle\tilde{A}_{k_2}^{\dagger}\tilde{A}_{k_2}\rangle$. The
equations of motion cast into
\begin{eqnarray}
i\frac{d}{dt}\tilde{\mathcal{A}}_{k_1}&=&\left(\Omega_{k_1}-\omega_1-i\Gamma_{k_1}+V_{k_1,k_2}\mathcal{N}_{k_2}\right)\tilde{\mathcal{A}}_{k_1}+\tilde{F}_{k_1}, \nonumber \\
i\frac{d}{dt}\tilde{\mathcal{A}}_{k_2}&=&\left(\Omega_{k_2}-\omega_2-i\Gamma_{k_2}+V_{k_1,k_2}\mathcal{N}_{k_1}\right)\tilde{\mathcal{A}}_{k_2}+\tilde{F}_{k_2}. \nonumber \\
\end{eqnarray}
The steady state solutions are obtained at
$i\frac{d}{dt}\tilde{\mathcal{A}}_{k_1}=i\frac{d}{dt}\tilde{\mathcal{A}}_{k_2}=0$. Hence the steady state average number of polaritons, in the mean field theory, can be defined by
$\mathcal{N}_{k_1}^{st}=|\tilde{\mathcal{A}}_{k_1}^{st}|^2$ and
$\mathcal{N}_{k_2}^{st}=|\tilde{\mathcal{A}}_{k_2}^{st}|^2$, and that are given by
\begin{eqnarray}
\mathcal{N}_{k_1}^{st}&=&\frac{|\tilde{F}_{k_1}|^2}{\left(\omega_1-\Omega_{k_1}-V_{k_1,k_2}\mathcal{N}_{k_2}^{st}\right)^2+\Gamma_{k_1}^2}, \\ \nonumber
\mathcal{N}_{k_2}^{st}&=&\frac{|\tilde{F}_{k_2}|^2}{\left(\omega_2-\Omega_{k_2}-V_{k_1,k_2}\mathcal{N}_{k_1}^{st}\right)^2+\Gamma_{k_2}^2},
\end{eqnarray}
and which can be solved for $\mathcal{N}_{k_1}^{st}$ and
$\mathcal{N}_{k_2}^{st}$.

We treat now pump-probe experiments in which one of the fields is taken to be strong and in resonance
where $\omega_1=\Omega_{k_1}$ and that
called the pump field, say the one with $k_1$, and results in a fixed
$\mathcal{N}_{k_1}^{pump}$. The second field with $k_2$ is the weak probe field
of $\tilde{F}_{k_2}^{probe}$, and that oscillates in frequency $\omega_2=\omega$, with the average number of polaritons
\begin{equation}
\mathcal{N}_{k_2}^{st}=\frac{|\tilde{F}_{k_2}^{probe}|^2}{\left(\omega-\Omega_{k_2}-V_{k_1,k_2}\mathcal{N}_{k_1}^{pump}\right)^2+\Gamma_{k_2}^2}.
\end{equation}
The spectrum can be obtained by changing $\omega$. The observed spectrum is a tool for evaluating the interaction strength. The blue shift in the probe spectrum is proportional to the
interaction strength at a fixed pump.

We plot the probe spectrum for the cases with and without a pump, in which
$\mathcal{N}_{k_2}^{st}(0)=\frac{|\tilde{F}_{k_2}^{probe}|^2}{\left(\omega-\Omega_{k_2}\right)^2+\Gamma_{k_2}^2}$. We
use the previous numbers, and for the excited atom damping rate we have
$\hbar\Gamma_A=2.15\times 10^{-8}\ eV$, and for the cavity photon damping rate
we take
$\hbar\Gamma_C=2.15\times 10^{-10}\ eV$. The average number of pump polaritons
is taken to be $\mathcal{N}_{k_1}^{pump}=100$. We plot the scaled
$\mathcal{N}_{k_2}^{st}/|\tilde{F}_{k_2}^{probe}|^2$ for arbitrary units as a
function of the detuning $\omega-\Omega_{k_2}$, at $k_2=10^{-6}\
\AA^{-1}$. Figure (6.a) is for the zero excitation-photon detuning case with $E_C(k)-E_A=0$, and figure (6.b) for
the detuning of $E_C(k)-E_A=-10^{-4}\ eV$. The shift in the probe spectrum
after switching on the pump field can be used to extract the strength of the
interaction for given average number of pump polaritons. It is clear that
for larger negative detuning the shift is much smaller than the zero
detuning case, as here the polaritons are much more photonic and weakly
interact. Moreover, the scattering through the kinematic interaction significantly reduce
the probe polariton intensity at the two cases.

\begin{figure}
\centerline{\epsfxsize=4.5cm \epsfbox{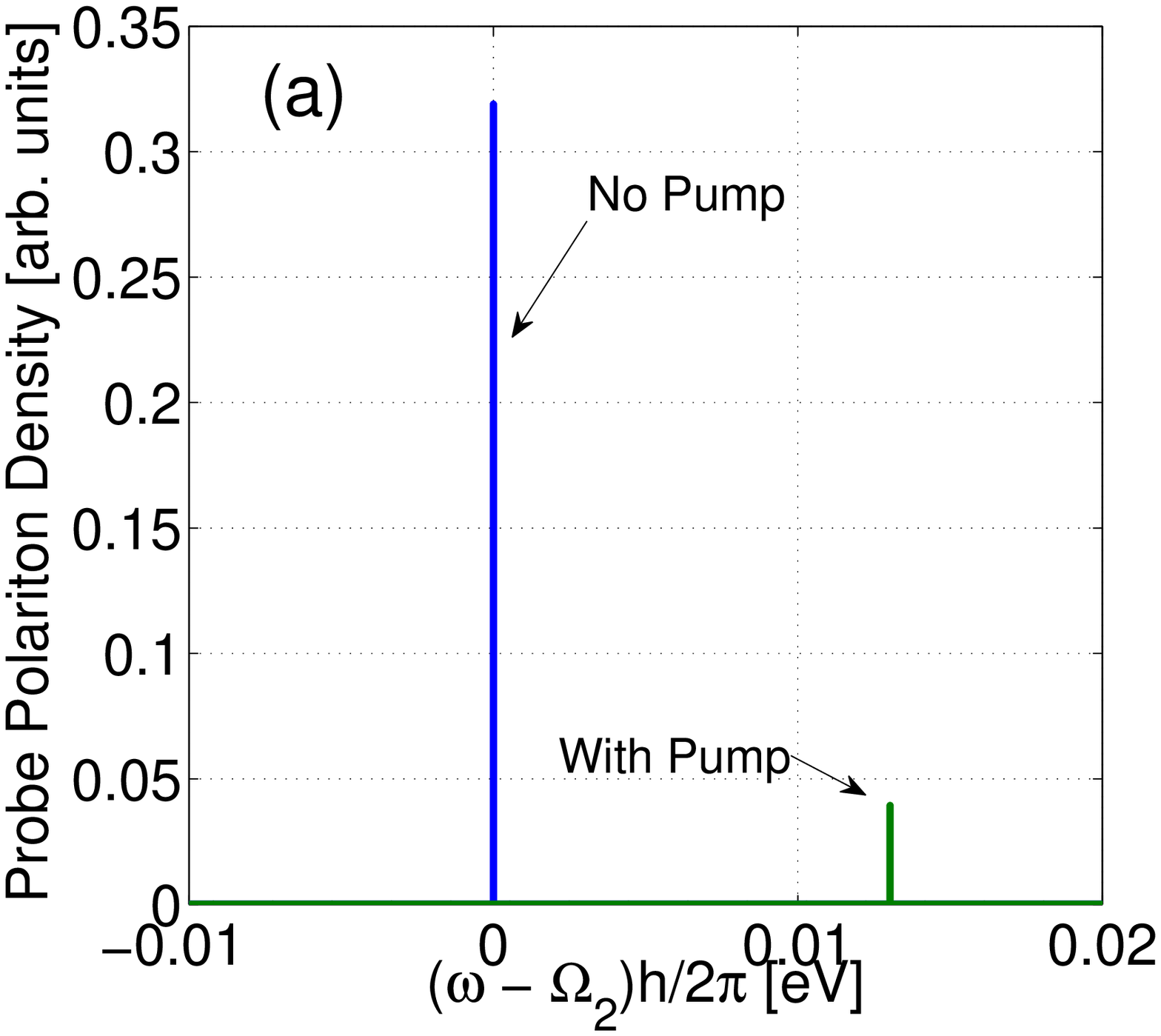}\epsfxsize=4.5cm \epsfbox{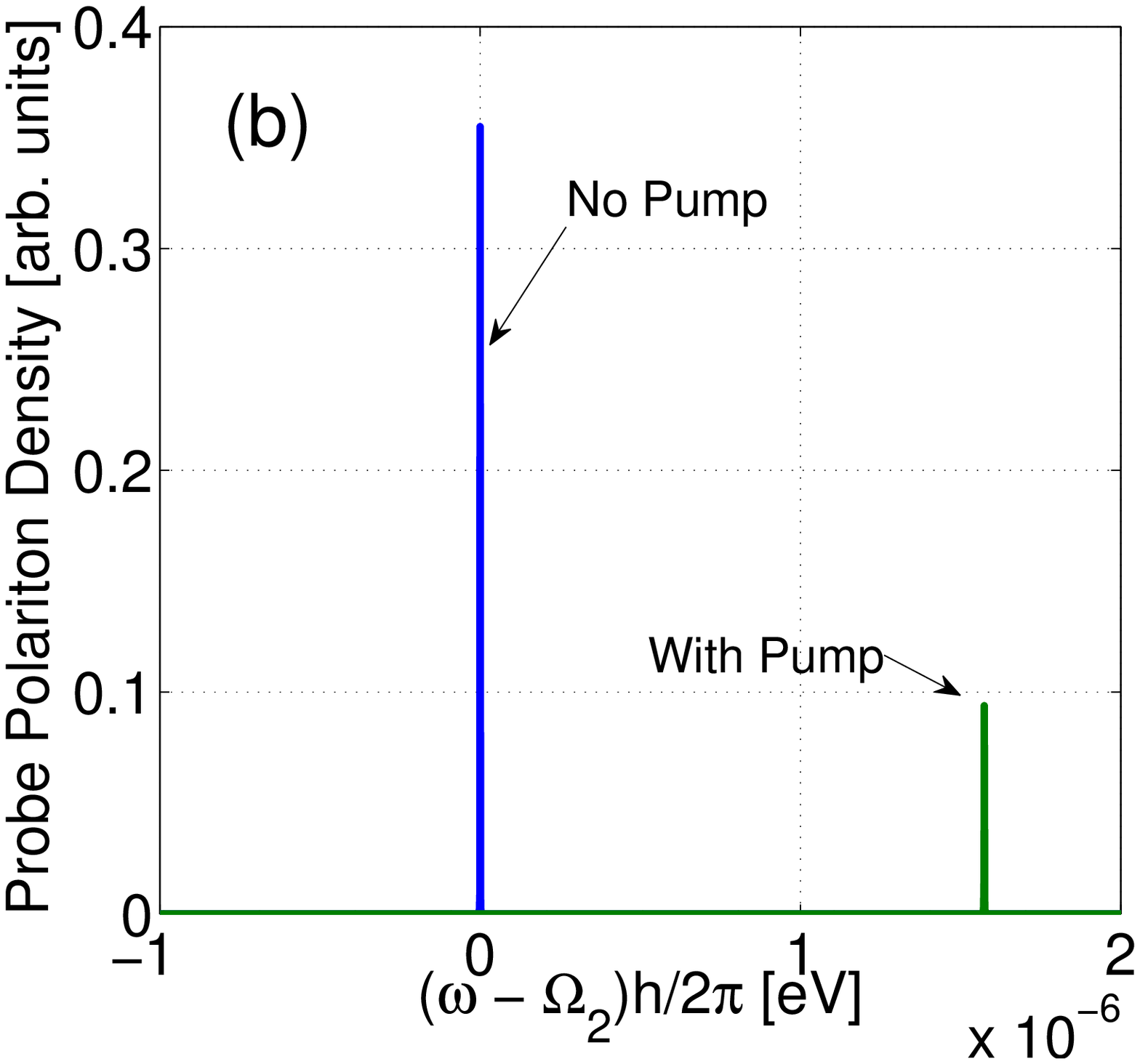}}
\caption{(a) The density of probe polaritons, with arbitrary units, as a function
  of the detuning  $\omega-\Omega_{k_2}$, for $k_2=10^{-6}\
\AA^{-1}$. (a) At zero atom-photon detuning, that is $E_C(k)-E_A=0$. (b) At
finite atom-photon detuning of $E_C(k)-E_A=-10^{-4}\ eV$.}
\end{figure}

Now we consider the case of two identical counter
propagating resonance external fields of
$\tilde{F}_{k_1}=\tilde{F}_{k_2}=\tilde{F}$, where $\omega_1=\Omega_{k_1}$ and
$\omega_2=\Omega_{k_2}$, with $\Omega_{k_1}=\Omega_{k_2}=\Omega$ and
$\Gamma_{k_1}=\Gamma_{k_2}=\Gamma$. We have $k_1=+k$ and $k_1=-k$. This fact
leads to a fixed $\mathcal{N}_{k_1}=\mathcal{N}_{k_2}=\mathcal{N}$. At steady state
$\tilde{\mathcal{A}}_{+k}^{st}=\tilde{\mathcal{A}}_{-k}^{st}=\frac{\tilde{F}}{i\Gamma-V\mathcal{N}}$,
where $V=V_{k_1,k_2}$. We have
$\mathcal{N}=|\tilde{\mathcal{A}}_{+k}^{st}|^2=|\tilde{\mathcal{A}}_{-k}^{st}|^2=\tilde{\mathcal{A}}_{+k}^{st\star}\tilde{\mathcal{A}}_{-k}^{st}=\tilde{\mathcal{A}}_{-k}^{st\star}\tilde{\mathcal{A}}_{+k}^{st}$, with
\begin{equation}
\mathcal{N}(\Gamma^2+V^2\mathcal{N}^2)=|\tilde{F}|^2,
\end{equation}
that needs to be solved for $\mathcal{N}$. In
figure (7) we plot the average polariton number $\mathcal{N}$ as a function of
the scaled external field intensity $|\tilde{F}|^2$ in using the previous
numbers, and for zero excitation-photon detuning, that
is $E_C(k)-E_A=0$, at $k=10^{-6}\
\AA^{-1}$. The finite detuning case of $E_C(k)-E_A=-10^{-5}\ eV$ is also
plotted. It is easier to achieve higher average number of polaritons at fixed external field for the case of finite detuning than in the case of zero detuning.

\begin{figure}
\centerline{\epsfxsize=6cm \epsfbox{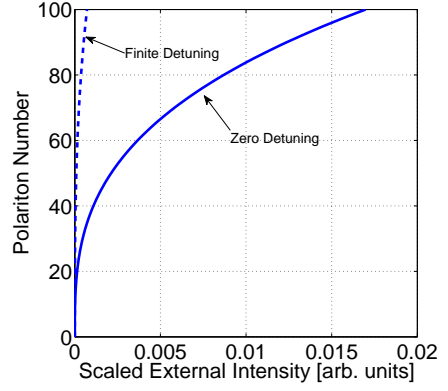}}
\caption{The average polariton number $\mathcal{N}$ vs. the scaled external field $|\tilde{F}|^2$, at zero atom-photon detuning, $E_C(k)-E_A=0$, and with $k=10^{-6}\
\AA^{-1}$. The dashed line is for the case of finite detuning with $E_C(k)-E_A=-10^{-5}\ eV$.}
\end{figure}

At this point we aim to calculate the correlation among two atoms setting at different
sites in the lattice, which are $(n)$ and $(m)$. Namely, we target to
calculate $\langle B_m^{\dagger}B_n\rangle$. But first we present the
excitation operator in terms of polariton ones by
$B_n=\frac{1}{\sqrt{N}}\sum_{k,r}X_k^re^{-ikz_n}A_k^r$. For the present case we have
\begin{equation}
B_n=\frac{1}{\sqrt{N}}\left\{X_{+k}e^{-ikz_n}A_{+k}+X_{-k}e^{ikz_n}A_{-k}\right\}.
\end{equation}
We have also $X_{+k}=X_{-k}=X$, and we get
\begin{eqnarray}
\langle B_m^{\dagger}B_n\rangle&=&\frac{|X|^2}{N}\left\{\langle
  A_{+k}^{\dagger}A_{+k}\rangle e^{-ik(z_n-z_m)}\right.  \\ \nonumber
&+&\left.\langle
  A_{-k}^{\dagger}A_{-k}\rangle e^{ik(z_n-z_m)}
+\langle A_{+k}^{\dagger}A_{-k}\rangle e^{ik(z_n+z_m)}\right.  \\ \nonumber
&+&\left.\langle A_{-k}^{\dagger}A_{+k}\rangle e^{-ik(z_n+z_m)}\right\}.
\end{eqnarray}
In the mean field theory and at steady state we obtain
\begin{equation}
\langle B_m^{\dagger}B_n\rangle^{st}_{mean}=\frac{4|X|^2\mathcal{N}}{N}\ \cos(kz_n)\cos(kz_m),
\end{equation}
where $\mathcal{N}$ is the solution of the above equation. If we fix the $(n)$ atom, e.g., at the origin with $z_n=0$
then the correlation have $cos$ behavior with $z_m$. The result emphasizes the
fact that photons mediate long range interactions among atoms.

Here we present the possibility of optical bistability behavior in such a system. We consider, as before, the case of two identical counter
propagating external fields of $\omega_1=\omega_2=\omega$. But most
important now
the external field is off resonance with the polaritons, and we have finite $\bar{\delta}=\omega-\Omega$. We conclude
that at steady state
\begin{equation}
\mathcal{N}\left[(\bar{\delta}-V\mathcal{N})^2+\Gamma^2\right]=|\tilde{F}|^2,
\end{equation}
and that gives $\mathcal{N}$. In
figure (8) we plot the average polariton number $\mathcal{N}$ as a function of
the scaled external field intensity $|\tilde{F}|^2$ in using the previous
numbers, with the external field-polariton detuning of $\bar{\delta}=10^{-3}\ eV$, at $k=10^{-6}\ \AA^{-1}$. We present the two cases of zero and finite
excitation-photon detunings, which are $E_C(k)-E_A=0$ and
$E_C(k)-E_A=-10^{-5}\ eV$. The plot show clear optical bistability
behavior. The optical bistability zone increases with
increasing the excitation-photon detuning.

\begin{figure}
\centerline{\epsfxsize=6cm \epsfbox{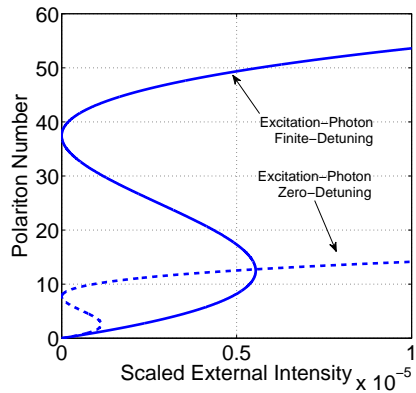}}
\caption{The average polariton number $\mathcal{N}$ vs. the scaled external
  field $|\tilde{F}|^2$, for the external field-polariton detuning of $\bar{\delta}=10^{-3}\ eV$, and with $k=10^{-6}\
\AA^{-1}$. The full line is at zero atom-photon detuning, $E_C(k)-E_A=0$, and
the dashed line is for the case of finite detuning with $E_C(k)-E_A=-10^{-5}\ eV$.}
\end{figure}

\section{Conclusions}

The present paper is opened by introducing polaritons as natural excitations
in a lattice of cold atoms strongly coupled to a nanophotonic waveguide, where a tapered
nanofiber is used as a prototype system. Afterward we exploited a bosonization procedure that end in
interacting polaritons by converting spin-half operators into combinations of boson ones with terms that result in polariton interactions. The
kinematic polariton-polariton interaction is introduced as a significant
mechanism for different nonlinear processes. The regime where polaritons
behave as a dilute boson gas is
presented, which is a step toward the possibility of achieving BEC of
polaritons in such system. Emphasize is put on lower branch
polaritons with long wavelengths that behave as quasi-particles with
effective mass and finite zero energy. The coherent mixing of dispersion-less
excitations and confined photons of parabolic dispersion leads to the above
features of polaritons. The excitation-photon detuning is presented as a tool
for controlling the kinematic polariton-polariton interaction strength.

The polariton-polariton scattering, due to the kinematic interaction, is for the
polariton excitation part that modeled in the center of mass frame as a
scattering of a polariton with a reduced mass from a potential that results of the
other polariton, and which appears as a defect in the
lattice. The polariton
excitation part is completely backward scattered, and we concluded that the polariton backward scattering have reflection
amplitude as their excitation amplitude, and the polariton forward scattering have transmission amplitude as their photonic
amplitude. Significant exception
appeared for large negative detuning, as here lower branch polaritons of long
wave lengths are much more photonic than excitation and they are weakly
interact via the kinematic interactions. Polaritons in this regime can be treated
as interacting photons, where the issue of superfluidity of photons is
proposed for such a system.

Different optical nonlinear processes are treated and presented as a tool for
observing physical properties of the system. Here the finite life time of
the fiber photons and of the excited atoms are included and external source fields contained. In the mean field theory we derived equations of motion for the
polariton amplitudes and solved for the steady state case. The pump-probe experiment is introduced, where the weak probe spectrum
calculated under the existence of fixed strong pump. The blue shift in the probe
spectrum found to be proportional to the kinematic interaction
strength. Furthermore, for identical counter propagating pump fields, we derived
the steady state average number of photons for various parameters. In this case
we extracted the atom-atom correlation of two separated atoms in the lattice
that indicates possible significant long range interaction mediated by the cavity
photons. For finite external pump-polariton detuning we got optical
bistability behavior in the average number of polaritons as a function of the external
pump intensity.

The above scattering results are directly adopted in the appendix for the scattering of polaritons
from an impurity atom of different kind than the lattice atoms. The
language of polaritons is shown to provide direct conclusions concerning
the scattering problem, and more useful cases can be engineered. The scattering
amplitude can be easily changed by controlling the impurity level relative
to the lattice atoms in using external fields. The results of the present paper can be adapted for any
set-up of one dimensional optically active materials that are
strongly coupled to one dimensional propagating photons.

\ 

The author gratefully acknowledges helpful and fruitful discussions with Thomas
Pohl.

\appendix

\section{Polariton Scattering from an Atom Impurity in the Lattice}

The appearance of defects in the atomic lattice is unavoidable in real
experiments. In the present appendix we discuss the scattering of a polariton from an impurity atom of
different type in the lattice that is localized in the origin, as seen in figure (9). The defect
Hamiltonian is given by $H_{d}=\bar{E}\ B_0^{\dagger}B_0$, where $\bar{E}=E_d-E_A$, with $E_d$ is the impurity atom transition energy. Using the previous results
for the scattering of small wave number polaritons with parabolic
dispersion, one get the square scattering amplitude
\begin{equation}
|f(k)|^2=\frac{\frac{\bar{E}^2}{\Lambda^2}\left(\frac{\pi^2}{2ka}\right)^2|X_{k}^{-}|^4}{1+\frac{\bar{E}^2}{\Lambda^2}\left(\frac{\pi^2}{2ka}\right)^2|X_{k}^{-}|^4}.
\end{equation}

\begin{figure}
\centerline{\epsfxsize=6cm \epsfbox{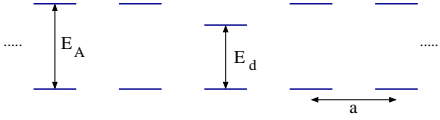}}
\caption{The atomic lattice of lattice constant $a$ with localized impurity of
  different transition energy $E_d$.}
\end{figure}

We present the results for the previous numbers. First we plot the
eigen-energies in figure (10.a)
and the excitation and photon fractions in figure (10.b) for the two branches as a function of
the excitation-photon detuning, $E_C(k)-E_A$, for fixed small wave number of
$k\approx 10^{-6}\
\AA^{-1}$, which is much smaller than $k_0$. Around zero
excitation-photon detuning the polariton
is half excitation and half photon. For large negative detuning the lower branch
becomes more photonic and the upper one more excitation, and for large positive
detuning the opposite the lower branch becomes more excitation and the upper
one more photonic. Note that the detuning need to be much smaller
than the atomic transition for the rotating wave approximation to hold.

\begin{figure}
\centerline{\epsfxsize=4.5cm \epsfbox{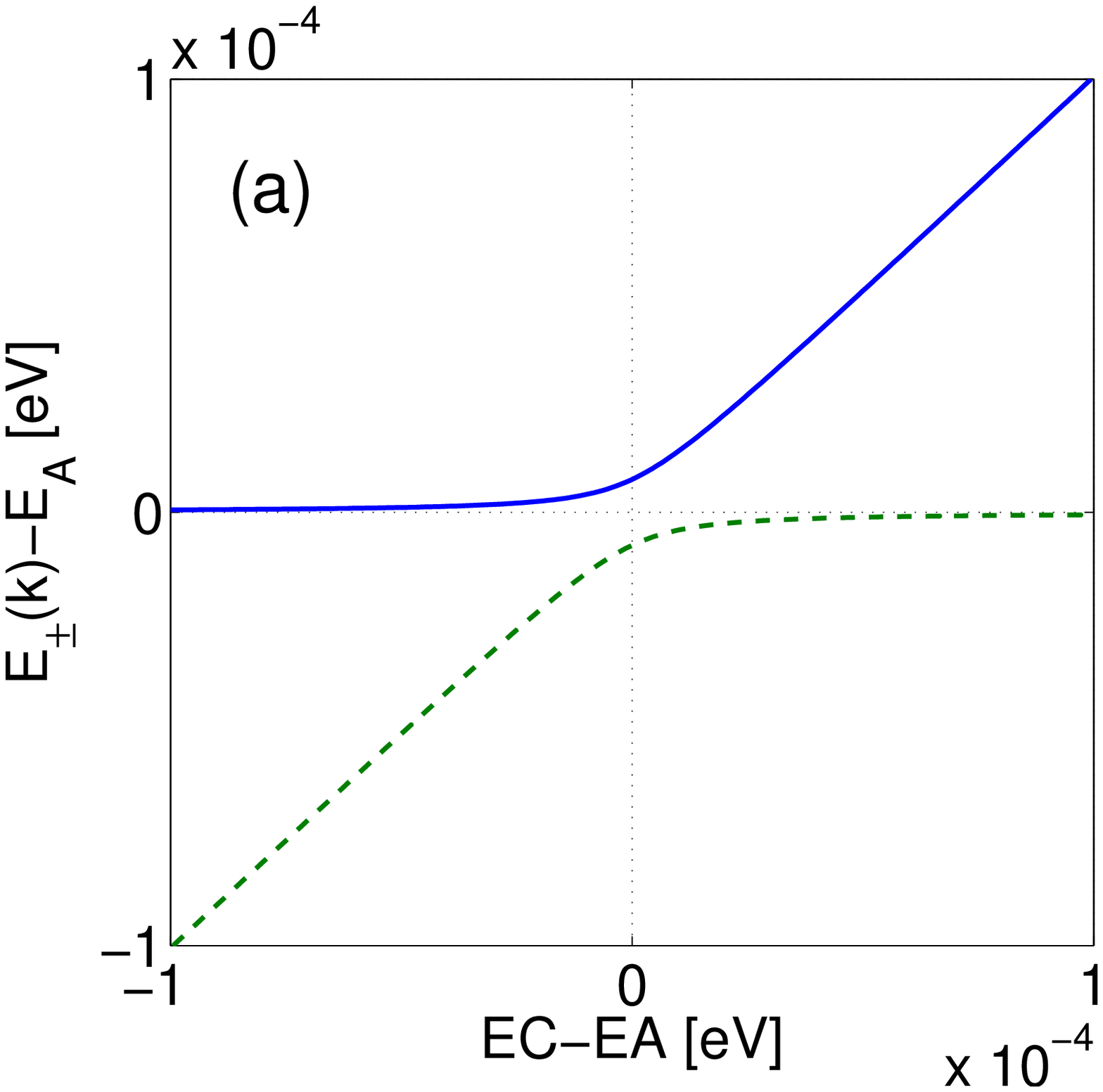}\epsfxsize=4.5cm \epsfbox{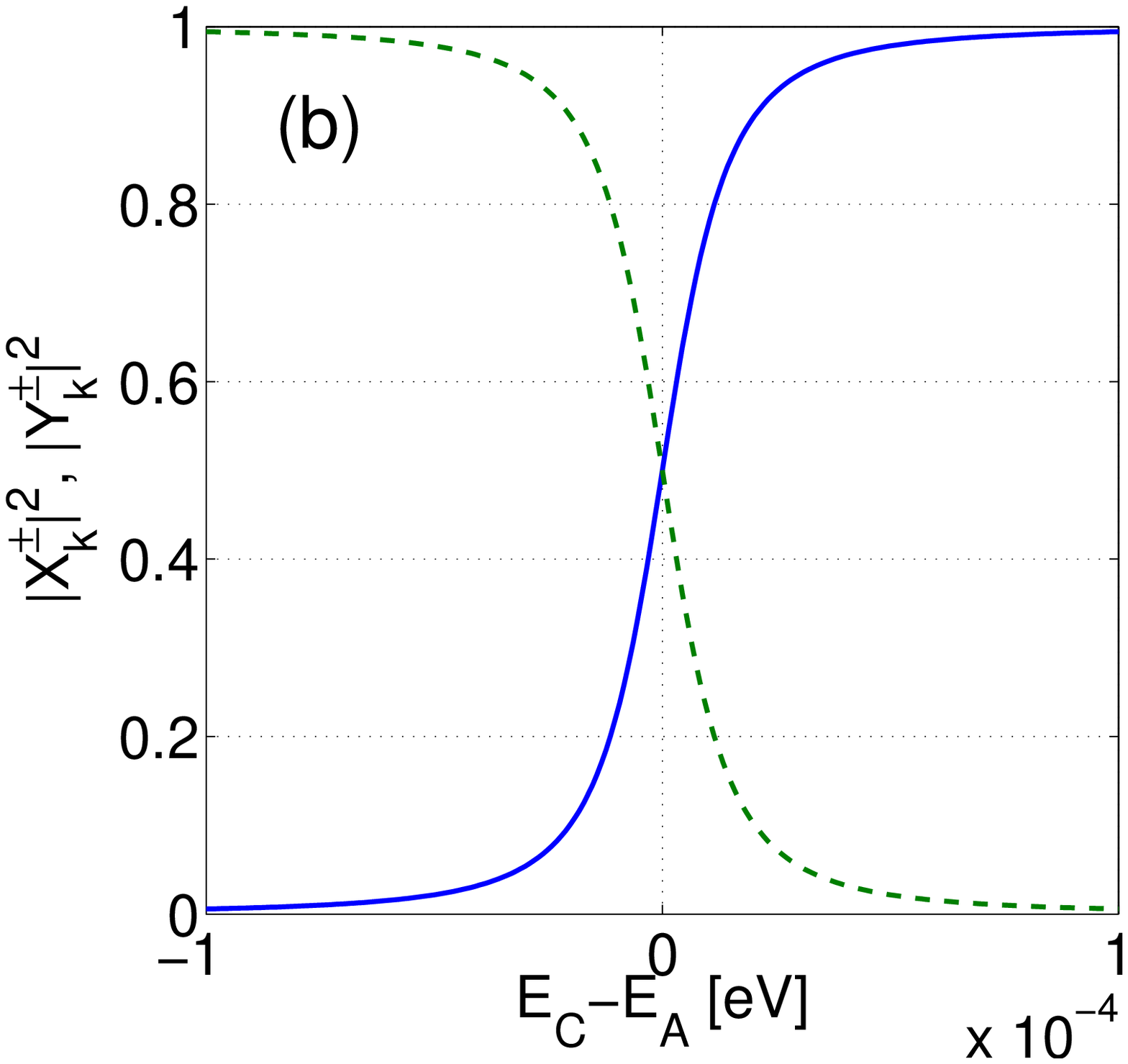}}
\caption{(a) The polariton relative energies $E_{\pm}(k)-E_A$ vs. the detuning $E_C(k)-E_A$ in the upper and
  lower branches by changing the photon energy, for $k=10^{-6}\ \AA^{-1}$. (b) The excitation and photon fractions in the two polariton branches
  vs. the detuning $E_C(k)-E_A$, for $k=10^{-6}\ \AA^{-1}$. In
  the upper branch the full line is for the photon fraction $|Y_k^{+}|^2$ and the
  dashed line for the excitation fraction $|X_k^{+}|^2$. In
  the lower branch the dashed line is for the photon fraction $|Y_k^{-}|^2$ and the
  full line for the excitation fraction $|X_k^{-}|^2$.}
\end{figure}

We plot in
figure (11) the
absolute square of the scattering amplitude, that is $|f(k)|^2$ as a function
of the detuning $E_C(k)-E_A$ for small fixed wave number of $k\approx 10^{-6}\
\AA^{-1}$, and for deep impurity with $E_d=0$. Things are drastically changed for
large negative detuning, as now the lower branch around small wave numbers
becomes more photonic and in the off-resonance case the scattering become
finite and drops down with increasing the negative detuning. 

\begin{figure}
\centerline{\epsfxsize=6cm \epsfbox{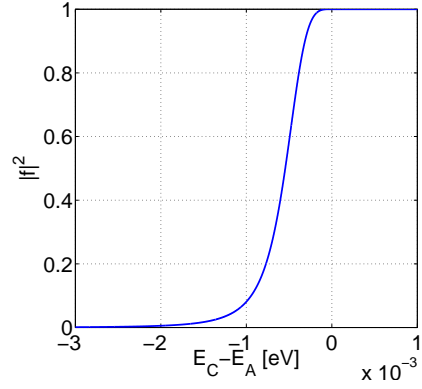}}
\caption{The absolute square of the scattering amplitude, $|f(k)|^2$ vs. the detuning
$E_C(k)-E_A$, for $k=10^{-6}\ \AA^{-1}$, and with $E_d=0$.}
\end{figure}

The polariton for large negative detuning is weakly scattered in the limit of
$\frac{\bar{E}}{\Lambda}\left(\frac{\pi^2}{2ka}\right)|X_{k}^{-}|^2\ll 1$, where the absolute value of the scattering amplitude is
\begin{equation}
|f(k)|\approx\frac{\bar{E}}{\Lambda}\left(\frac{\pi^2}{2ka}\right)|X_{k}^{-}|^2.
\end{equation}
In the limit of large negative detuning the parabolic assumption holds and the
contribution of the upper branch is negligible. Here the polaritons
are weakly scattered from the vacancy.

\begin{figure}
\centerline{\epsfxsize=6cm \epsfbox{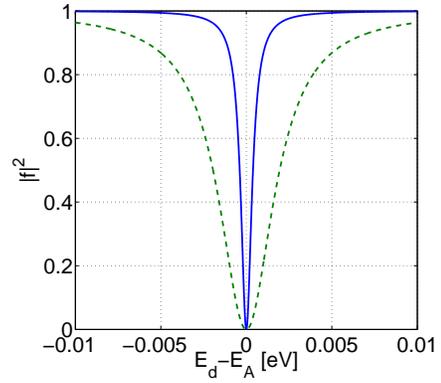}}
\caption{The absolute square of the scattering amplitude, $|f(k)|^2$ vs. the detuning
$E_d-E_A$, for (full-line) $ka=5\times 10^{-3}$, and for (dashed-line) $ka=25\times 10^{-3}$.}
\end{figure}

As the defect potential is relatively a deep square well potential of
depth $E_A$ and width $a$, the
scattering is the same as that from a square barrier potential of height $E_A$
and width $a$. Namely, we have the known result that for deep potentials the appearance
of bound states is immaterial as the probability to hit a resonance is very
small. Hence the defect is repulsive potential where the polariton excitation
part backward reflected, and only the polariton photon part have forward
scattering or move through the defect without interaction.

For the case of $E_C(0)-E_A=0$, we have
\begin{equation}
|f(k)|^2=\frac{\frac{\bar{E}^2}{\Lambda^2}\left(\frac{\pi^2}{4ka}\right)^2}{1+\frac{\bar{E}^2}{\Lambda^2}\left(\frac{\pi^2}{4ka}\right)^2}.
\end{equation}
In figure (12) we plot the  square scattering amplitude, $|f|^2$, as a
function of $\bar{E}$, for small wave number polaritons of $ka=5\times
10^{-3}$, and $ka=25\times 10^{-3}$, and
in using the previous numbers where $\Lambda\approx 0.19\ eV$. It is clear
that scattering resonance appears at $\bar{E}=0$ where the impurity become
similar to the other lattice atoms. Complete scattering
appears at large positive and negative $\bar{E}$. The resonance width is
related on the wave number, where it is wide for large $k$ and narrow for
small one. For the case of large positive and negative $\bar{E}$ the polariton
transmission is $|Y_0|^2$ and the polariton reflection is $|X_0|^2$, which are
$1/2$ for the case of $E_C(0)-E_A=0$.

The localization of two identical impurities at separation
distance that is larger than $a$ gives rise to atomic mirror, as in the paper \cite{Chang},
but here the results can be obtained directly in terms of polaritons. The
scattering of polaritons from a defect in planar optical lattice in the Mott
insulator phase with one atom per site has been studied in \cite{ZoubiE,ZoubiRev}.

\end{document}